\documentclass[%
  reprint,
 amsmath,amssymb,
 aps,
]{revtex4-1}
\usepackage{graphicx}
\usepackage{dcolumn}
\usepackage{bm}
\usepackage{xcolor}
\usepackage{latexsym}
\usepackage{comment}

\begin{document}

\title{Balanced--imbalanced transitions in indirect reciprocity dynamics on networks}

\author{Koji Oishi}
\email{oishi@sipeb.aoyama.ac.jp}
\affiliation{Department of International Politics, Aoyama Gakuin University, Tokyo, Japan}
\affiliation{Japan Society for the Promotion of Science, 5-3-1 Kojimachi, Chiyoda-ku, Tokyo 102-8471, Japan}

\author{Shuhei Miyano}
\affiliation{Department of Applied Physics, Graduate School of Engineering, The University of Tokyo, 7-3-1 Hongo, Bunkyo-ku, Tokyo, 113-8656, Japan}

\author{Kimmo Kaski}
\email{shimada@sys.t.u-tokyo.ac.jp}
\affiliation{Department of Computer Science, Aalto University School of Science, PO Box 15500, Espoo, Finland}
\affiliation{The Alan Turing Institute, British Library, 96 Euston Road, London NW1 2DB, UK}

\author{Takashi Shimada}
\email{shimada@sys.t.u-tokyo.ac.jp}
\affiliation{Mathematics and Informatics Center}
\affiliation{Department of Systems Innovation, Graduate School of Engineering, The University of Tokyo, 7-3-1 Hongo, Bunkyo-ku, Tokyo 113-8656, Japan.}

\begin{abstract}
  Here we investigate the dynamics of indirect reciprocity on networks, a type of social dynamics in which the attitude of individuals, either cooperative or antagonistic, toward other individuals changes over time by their actions and mutual monitoring. We observe an absorbing state phase transition as we change the network's link or edge density. When the edge density is either small or large enough, opinions quickly reach an absorbing state, from which opinions never change anymore once reached. In contrast, if the edge density is in the middle range the absorbing state is not reached and the state keeps changing thus being active. 
  The result shows a novel effect of social networks on spontaneous group formation.
\end{abstract}

\maketitle

\section{Introduction}
Social relations are characterized by being of
either cooperative (friendly) or antagonistic (unfriendly) nature. However, it can also happen that friends turn to foes or foes to friends~\cite{Harrigan2020}, i.e. an individual network link turn from positive to negative or negative to positive. Hence understanding the temporal evolution of such behaviour changing processes is essential for getting deeper insight into the functions of real social networks. 

In order to gain such insight, Agent Based Modelling approaches have turned out to be versatile. One of the earlier models on how people change their attitude towards others was based on indirect reciprocity~\cite{Nowak1998,Ohtsuki2004}, which is commonly observed in human behavior and is closely related to the evolution of cooperation in human society~\cite{Nowak2006}.
In this model, people change their attitude (of either liking or disliking) through their action and observation of the actions of others.
A typical example of such dynamics (action rules and norms) is as the following.
First, people are cooperating with only those they like.
Second, they get to like those who cooperate with those whom they like and do not cooperate with those whom they dislike.

In spite of the importance of indirect reciprocity, the dynamics of social networks induced by indirect reciprocity has not yet been fully understood.
Previous studies have focused on the case of fully connected networks (i.e., systems in which all people know each other)~\cite{Oishi2013,Isagozawa2016}.
It was found that indirect reciprocity can result in a split of the society into clusters, only within which people are cooperative.
However, the real social networks are usually not fully connected. Therefore, we focus our attention on the dynamics of indirect reciprocity in such networks of agents and examine whether agents split into fixed clusters as in the case of fully connected networks.
In this study, among several variations of the indirect reciprocity dynamics, we focus on the Kandori assessment rule~\cite{Kandori1992}. Comparing to another type of indirect reciprocity model in which an agent can get to like those who cooperate with those who he dislikes, the Kandori rule is said to be more strict because an agent dislikes the people in the same case. And this rule is suggested to be one of the most efficient and robust rules to promote cooperation~\cite{Kandori1992, Ohtsuki2004}.

This paper is organised such that after this Introduction, in Section~\ref{sec:model} we describe the model of indirect reciprocity.
Then we first present the simulation results in Section~\ref{sec:simulation} and then develop the mean field analysis for the model in~\ref{sec:mean-field}.
Finally in Section~\ref{sec:discussion} we summarize the results and discuss their implications. 

\section{The model}\label{sec:model}
\begin{figure}[!htb]
  \begin{center}
    \includegraphics[width=9cm]{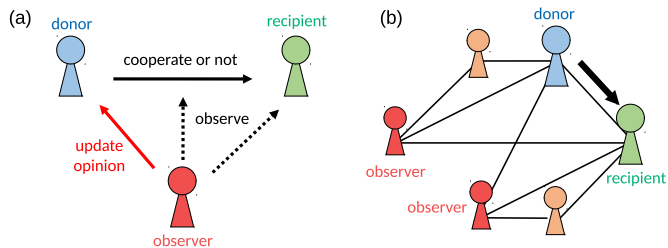}
    \caption{Schematics of the model.
      (a) Donor, recipient, and a third-party observer.
      (b) Observers in a network.
      Agents not labeled either donor, recipient, or observers do not observe the interaction between the donor and recipient.
    }
    \label{fig:model_schematics}
  \end{center}
\end{figure}

Let us consider a non-directed network of agents $G = (V, E)$, where $V = {1, \cdots, N}$ is the set of agents or network nodes and $E$ the set of links or network edges $e_{ij} \in E$ between agents $i$ and $j$, meaning that they know each other. Here the structure of the network is assumed not to change in time i.e. being static, while the agents have their opinion about their neighbors and change it in time i.e. being dynamic.
We denote the opinion of the agent $i$ about the agent $j$ by $\sigma_{ij}$.
If the agent $i$ likes the agent $j$ at time step $t$ then $\sigma_{ij}(t) = 1$, but if the agent $i$ dislikes $j$ then $\sigma_{ij}(t) = -1$.
The opinions do not need to be reciprocal so that $i$ may dislike $j$ even if $j$ likes $i$.

The time evolution of the model is set in such a way that at each time step, two neighboring agents are randomly chosen, one as the donor and the other as the recipient. The donor cooperates with the recipient if the donor likes the recipient, while the donor does not cooperate if the donor dislikes the recipient. The action of the donor, either cooperating or not cooperating,
is observed by the donor (him- or herself), the recipient, and the common neighbors of the donor and the recipient, i.e. the third party called the observer, as depicted in Fig. \ref{fig:model_schematics} (a).
Each observer of the donor-recipient pair updates her/his opinion about the donor according to a given rule, which is called the assessment rule.

In this study, we adopt the Kandori assessment rule~\cite{Kandori1992}:
observers get to like the donor if the observers like the recipient and the donor cooperates with the recipient or when observers dislike the recipient and the donor does not cooperate, while the observers get to dislike the donor otherwise.
Therefore, the opinion of an observer $k$ about the donor is updated as follows
\begin{equation}
  \sigma_{kd}(t+1) = \sigma_{kr}(t)\sigma_{dr}(t), \label{eq:update}
\end{equation}
where $d$ and $r$ denote the donor and the recipient at time step $t$, respectively.
Agents other than the 
donor, recipient, and the 
observers do not update their opinions (Fig.~\ref{fig:model_schematics} (b)). The 
agents' initial opinions are drawn independently at random from an even distribution of opinions, i.e. $\{-1,+1\}$, where $-1$ corresponds to antagonistic or unfriendly and $+1$ cooperative or friendly opinion (i.e., liking or disliking, respectively).

\section{Numerical results}\label{sec:simulation}
In this section we investigate the dynamics of indirect reciprocity based on the Kandori assessment rule on Erd\"{o}s--R\'{e}nyi random graphs, where each agent is linked to another agent with the probability $p$.
In the following, the ensemble averages are taken over independently generated networks.
Note that the number of links can be different across samples even with the same probability $p$, due to stochastic fluctuations.

\begin{figure}[bt]
  \begin{center}
    \includegraphics[width=8cm]{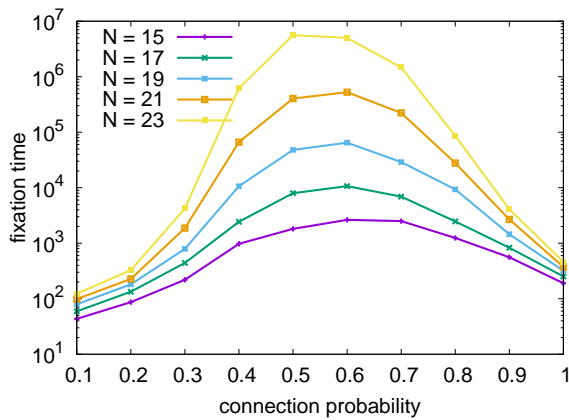}
    \caption{Average fixation time. We conduct $100$ independent runs and take the average over samples that have reached to absorbing states by $10^{7}$ time steps.}
    \label{fig:fixtime}
  \end{center}
\end{figure}

First we focus on the question whether the opinions of agents eventually becomes fixed, as was found in the case of fully-connected networks~\cite{Oishi2013}.
Following the update rule (i.e., Eq.~\ref{eq:update}), the condition that opinions do not change anymore is
\begin{equation}
  \sigma_{kd} = \sigma_{kr}\sigma_{dr},
\end{equation}
for all the network edges $e_{dr} \in E$ and all the observers $k$, i.e. $d$, $r$, and common neighbors of $d$ and $r$.
This condition is equivalent with
\begin{align}
  \sigma_{ii} &= 1 &\text{for all} ~ i \in V \label{eq:node_cond}\\
  \Theta_{ij} &\equiv \sigma_{ij}\sigma_{ji} = 1 &\text{for all} ~ e_{ij} \in E \label{eq:edge_cond}\\
  \Phi_{ijk} &\equiv \sigma_{ij}\sigma_{ik}\sigma_{jk} = 1 &\text{for all triads}~(i,j,k), \label{eq:triad_cond}
\end{align}
where $\Theta_{ij}$ and $\Phi_{ijk}$ stand for the edge balance and triad balance, respectively.
Because the system cannot move anymore after reaching a configuration which fulfills these conditions, we call all such configurations as absorbing state. And because the system always has absorbing state (e.g. $\forall i,j \ \ \sigma_{ij} = 1$), the question is whether and when the system goes to absorbing state.
In Fig.~\ref{fig:fixtime} we depict the average fixation time, the number of time steps taken until Eqs. (\ref{eq:node_cond}), (\ref{eq:edge_cond}), and (\ref{eq:triad_cond}) are satisfied.
When the connection probability $p$ is high or low, the fixation time is relatively short, while the time for the opinions getting fixed or fixation time turns out to be much longer with $p$ in the middle range and it rapidly increases even with a small increase of the system size $N$. This means that if the network is not fully-connected and the system size or the number of agents is moderate, i.e. $N > 100$, their opinion cannot split into fixed clusters in a realistic time scale.

The next question is how the opinion of an agent evolves before it reaches the absorbing state.
To investigate the time development, we introduce an order parameter imbalance $\mathcal{I}$ as
\begin{equation}
  \mathcal{I} = \begin{cases}
    1 - \frac{\sum_{i,j,k} \Phi_{ijk}}{6N_{\text{triad}}} & (N_{\text{triad}} > 0)\\
    0 & (N_{\text{triad}} = 0),
  \end{cases}
  \label{eq:def_order_paramter}
\end{equation}
where $N_{\text{triad}}$ is the number of agent triads.

As we show here, with some reasonable assumptions, the imbalance is equal to zero in the absorbing states and expected to be 1 in the case of random opinions.
Therefore, the order parameter evaluates how far the system is from the absorbing state and how near to the random state.
Let us examine Eq.~(\ref{eq:update}) to see how the order parameter works.
First, the update of the self-image of the donor at time step $t$ is as follows 
\begin{equation}
  \sigma_{id}(t+1) = \left(\sigma_{dr}\left(t\right)\right)^{2} = 1. \label{eq:update_donor}
\end{equation}
It means that the agents like themselves and never change the opinion once they take an action as a donor. Therefore, the first condition (Eqs. (\ref{eq:node_cond})) is quickly satisfied for all the agents. Next, the update of the opinion of the recipient on the donor follows,
\begin{equation}
  \sigma_{rd}(t+1) = \sigma_{rr}(t)\sigma_{dr}(t).
\end{equation}
After some transient time, i.e. $t \gg 1$, where $\sigma_{rr} =1$ as we have just discussed above, it reduces to
\begin{equation}
  \sigma_{rd}(t+1) = \sigma_{dr}(t). \label{eq:update_isolated_edge}
\end{equation}
Then if the edge $e_{ij}$ does not have any common neighbors, once $i$ takes an action to $j$ or vice versa, then $\sigma_{ij} = \sigma_{ji}$ is realized and they never change their opinion.
On the other hand, if the edge $e_{ij}$ has common neighbors, the relation $\sigma_{ij} = \sigma_{ji}$ is not always maintained in the long run, because $\sigma_{ij}$ may change when the agent $i$ observe the agent $j$'s action to a common neighbor.
Therefore, a non--trivial condition for the absorbing states is,
\begin{equation}
  \sigma_{ij} = \sigma_{ji}, \label{eq:condition_reciprocity}
\end{equation}
for all the edges $e_{ij}$ having common neighbors.
Finally, the update of the opinion of common neighbors cannot reduce from Eq. (\ref{eq:update}), therefore another non-trivial condition for fixed state is
\begin{equation}
  \sigma_{ij}\sigma_{ki}\sigma_{kj} = 1. \label{eq:condition_balance}
\end{equation}
Summing up these arguments, $\mathcal{I} = 0$ is equivalent to the condition for the opinion being in a fixed state, under the assumption $\sigma_{ii} =1$ and $\sigma_{ij} = \sigma_{ji}$ for all the edges $e_{ij}$ not having common neighbors, which are quickly satisfied.

\begin{figure}[bt]
  \begin{center}
    \includegraphics[width=8cm]{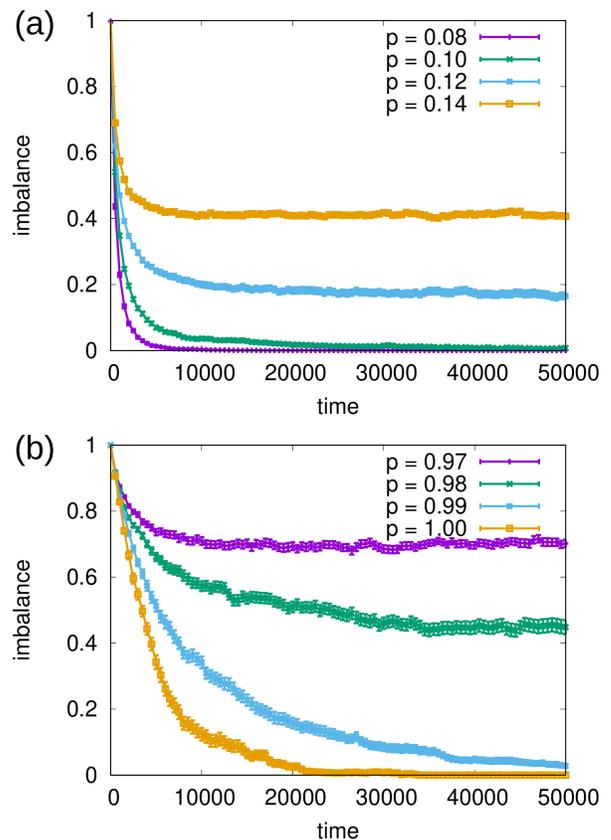}
    \caption{Time development of imbalance, averaged over 100 samples, in the case of (a) sparse and (b) dense random networks with $N=100$ agents. Error bars represent standard error.
    }
    \label{fig:imb_timeseries}
  \end{center}
\end{figure}

In Fig. \ref{fig:imb_timeseries} we show the time development of the imbalance for the network of size $N = 100$.
When the connection probability $p < 0.01$ or $p > 0.99$, the imbalance goes quickly down to zero.
In contrast, the imbalance remains positive if $0.01 < p < 0.99$.
This means that in the middle range values of the edge density, the system relaxes to a non-absorbed stationary state.

\begin{figure}[tb]
  \begin{center}
    \includegraphics[width=8cm]{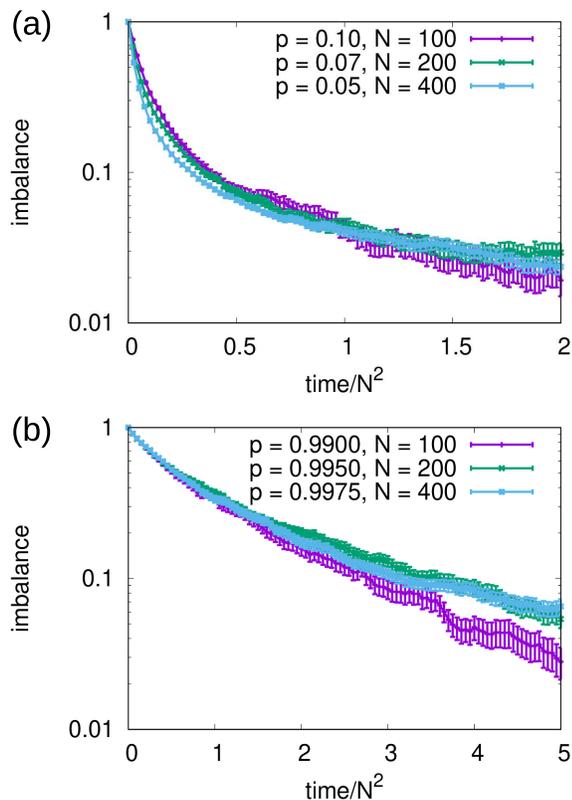}
    \caption{
      Imbalance as a function of time. The effect of the system size on the time development.
      The imbalance are averaged over 100 samples.
      Time series for different $p$ and $N$ are compared with (a) $pN^2 = 1$ and (b) $(1-p)N = 1$.
    }
    \label{fig:imb_scaling}
  \end{center}
\end{figure}

We also investigate the effect of the system size $N$ on the time development.
When the network is sparse (Fig. \ref{fig:imb_scaling} (a)), the time series of imbalance
collapses with the scaling of the time step $t$ to $t/N^2$ and the connection probability $p$ to $p/N^2$.
On the other hand, when the network is sparse (Fig. \ref{fig:imb_scaling} (b)), the time series are
scaled with $t/N^2$ and $p/(1-N)$. In both cases, the time development becomes $N^2$ times slower when the system size gets larger.

\begin{figure}[tb]
  \begin{center}
    \includegraphics[width=8cm]{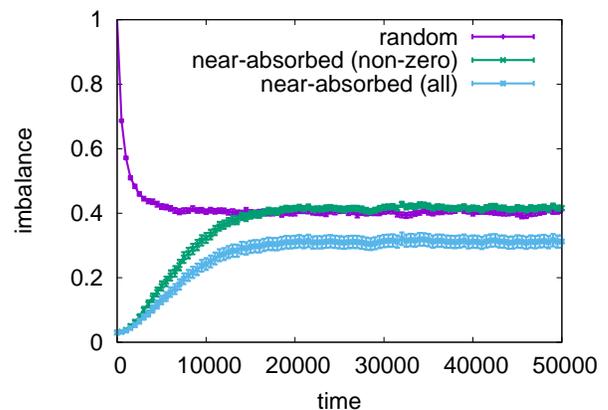}
    \caption{Average imbalance of the samples which starts from random initial condition (purple bars), average in the samples which starts from the vicinity of the absorbing state and are not absorbed in the observation period (green crosses), and the average in all the samples starting from the vicinity of absorbing state (cyan bars). The difference between the latter two describes the ratio of the absorbed samples. The average of the ``active'' samples starting from the vicinity of the absorbing state overtakes the one of the samples starting from the random initial condition, suggesting that there appears no hysteresis on this parameter and a small fraction of the systems starting from the random initial condition relaxes to the absorbing state.
    }
    \label{fig:imb_initial_condition}
  \end{center}
\end{figure}

We also confirm that the system has stationary states independent of the initial states.
We compare the results of the samples starting from the random initial state (our default setting) and those from the vicinity of absorbing states.
To generate the latter, we divide the agents into two groups of the same size and set them to like each other in the same groups and dislike those in the other groups with probability $0.99$, if they are connected. Opinions on the rest of the connection are drawn randomly and evenly to be either $-1$ or $+1$. Note that the group assignment is irrelevant with the network structure.
In Fig. \ref{fig:imb_initial_condition}, we compare the average time development of the imbalance.
If we average all samples, the stationary imbalance is smaller when the initial states are nearby the absorbing states.
However, this is because a certain fraction of samples get absorbed when starting nearby the absorbing states while samples starting from random opinions seldom get absorbed.
When we average only the samples that start nearby the  absorbing states and are not absorbed in each time step, the stationary imbalance is quite similar with the samples from the random initial states.

\begin{figure}[tb]
  \begin{center}
    \includegraphics[width=8.5cm]{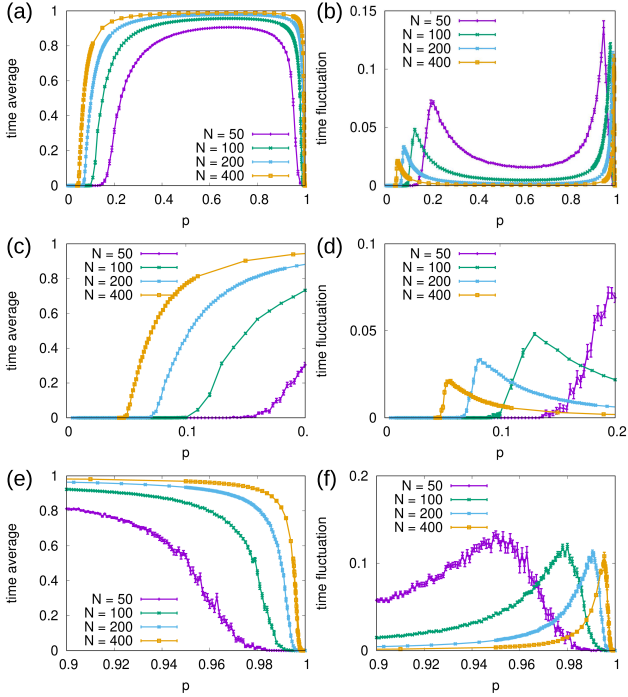}
    \caption{ (a, c, e) Time-average and (b, d, f) time-fluctuation of stationary imbalance on random networks.
      Time-average and time-fluctuation are calculated from 100 measurements from $5 \times N^2$ to $10 \times N^2$ time steps.
      Lines represent sample average and error bars for sample standard error over 100 different time-series for $N \leq 200$ and 50 time-series for $N = 400$.
      Panels (c, e) and (d, f) show the narrower part (around the transition points) of (a) and (c), respectively.
    }
    \label{fig:imb_stationary}
  \end{center}
\end{figure}

As the next step, we examine more closely how the stationary imbalance depends on the edge density.
In Fig. \ref{fig:imb_stationary} we show the time-average and time-fluctuation of imbalance after letting the network of agents to run for $5 \times N^2$ time steps of relaxation.
We observe that the stationary imbalances for the different system sizes
are scaled well by $Np^2$ in the sparsely connected networks and by $N(1-p)$ in the densely connected networks, as depicted in Figs. \ref{fig:imb_stationary} (c) and (e).
When $Np^2 < C_L \sim {\mathcal O}(1)$ or $N(1-p) > C_U \sim {\mathcal O}(1)$,
i.e., the network is quite sparse or quite dense, the stationary imbalance is equal to zero and the system reaches absorbing states, while both of them rapidly increase, the system remains more random with larger fluctuation as $Np^2(1-p)$ exceeds $1$.
The temporal fluctuations show peaks at around the same scaled boundary area.
This means that the systems remain far from absorbing states for the moderate edge density, while for more sparse or dense edge densities they show large fluctuations. However, when the edge density is quite sparse or dense and $p$ exceed certain values, the system reaches absorbing states.

These observations indicate that the edge density $p$ causes transitions between the absorbing phase in which the system goes to absorbing state and the active phase in which the system does not relax to the absorbing state \cite{Hinrichsen2000,Lubeck2004}.
Note that, as $N$ increases, the absorbing regions of $p$ (basins) get narrower: lower transition point $p_* \sim N^{-1/2}$ and higher transition point $p^* \sim 1 - 1/N$. We also note that more detailed analysis of the phase transition behavior with the order of the transition, accurate transition point, and possible critical exponents, is beyond the scope of this study. Instead we will next focus on the mean-field analysis.

\begin{figure}[tb]
  \begin{center}
    \includegraphics[width=8.5cm]{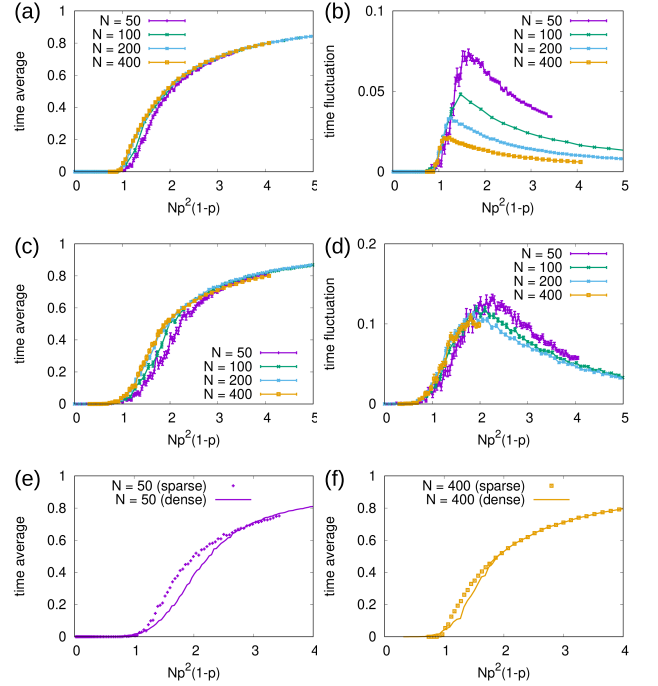}
    \caption{ Dependency of stationary imbalance on $Np^2(1-p)$.
      (a, c) Time-average and (b, d) time-fluctuation.
      (a, b) Sparse and (c, d) dense edge density regions.
      (e, f) Comparison of the time-average between sparse and dense edge density regions with (e) $N = 50$ and (f) $N = 400$.
    }
    \label{fig:imb_stationary_scaling}
  \end{center}
\end{figure}

\section{Mean-field analysis}\label{sec:mean-field}
\begin{figure}[tb]
  \includegraphics[width=3cm]{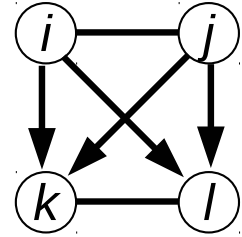}
  \caption{Tetrad balance.}
  \label{fig:quad_balance}
\end{figure}

In this section, we show that the transitions between the absorbing and active phases at $Np^2(1-p) \sim 1$ are consistent with a mean-field approximation for the indirect reciprocity dynamics.
For our analysis, we introduce tetrad balance (Fig. \ref{fig:quad_balance}) for each tetrad (four-node clique) as
\begin{equation}
    \Psi_{ijkl} \equiv
    \sigma_{ik} \sigma_{il} \sigma_{jk} \sigma_{jl}
    = \Phi_{ikl} \Phi_{jkl},
\end{equation}
in addition to the edge balance $\Theta_{ij}$ and triad balance $\Phi_{ijk}$ already defined in Eqs.~(\ref{eq:edge_cond}) and (\ref{eq:triad_cond}).
Note that the tetrad balance is by definition invariant under the exchange of the first two suffixes and the latter suffix pair:
\begin{equation}
    \Psi_{ijkl}
    = \Psi_{jikl}
    = \Psi_{jilk}.
\end{equation}

For the mean-field analysis, we use the averages of these quantities
\begin{eqnarray}
    \Theta &\equiv&
    \langle \Theta_{ij} \rangle
    = \frac{\sum_{i>j} \Theta_{ij} }{N_{\text{edge}}} \\
    \Phi &\equiv&
    \langle \Phi_{ijk} \rangle,
    = \frac{\sum_{i,j,k} \Phi_{ijk}}{6 N_{\text{triad}}}
    \qquad \Big( = 1 - \mathcal{I} \Big),
    \\
    \Psi &\equiv&
    \langle \Psi_{ijkl} \rangle
    = \frac{\sum_{i>j}\sum_{k>l} \Psi_{ijkl}}{6 N_{\text{tetrad}}},
\end{eqnarray}
as the order parameters of indirect reciprocity dynamics, where $N_{\text{edge}}$, $N_{\text{triad}}$, and $N_{\text{quad}}$ are the number of edges, triads, and tetrads in the system.

\begin{figure}[bt]
  \includegraphics[width=8cm]{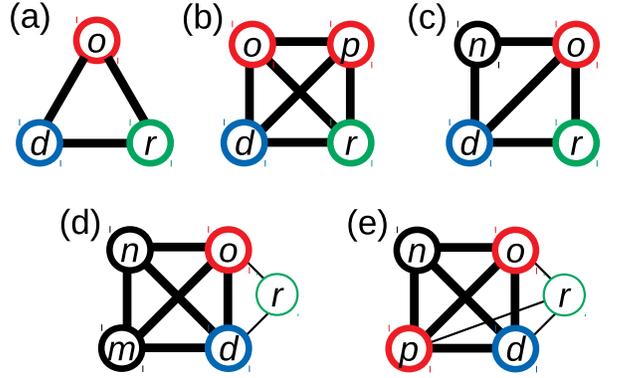}
  \caption{The subgraphs affected by the game between the donor $d$ and recipient $r$, which includes third-party observers $o$ and $p$, and non-observing neighbors $n$ and $m$.
    (a) Triads of $d$, $r$, and $o$.
    (b) Tetrads of $d$, $r$, $o$ and $p$.
    (c) Trusses of $d$, $r$, $o$, and $n$.
    (d) Tetrads of $d$, $o$, $n$ and $m$.
    (e) Tetrads of $d$, $o$, $p$ and $n$.
    Note that the recipient in (d) and (e) are shown only to illustrate the difference between third-party observers and non-observing neighbors, and is not included in the focal tetrads.
  }
  \label{fig:affected_subgraphs}
\end{figure}

As shown in the Appendix, a mean-field approximation on the indirect reciprocity dynamics yields the dynamical equations of the order parameters in the following closed form:
\begin{eqnarray}
    \frac{d \Theta}{dt} &=& \frac{1}{N_\text{edge}}\left[ (1 - \Theta) + T (\Phi - \Theta)\right],
    \label{eq:Theta_ODE}
    \\
    \frac{d \Phi}{dt}
    &=& \frac{1}{6N_\text{triad}}\left[ T \left\{ 2(1 -\Phi) + (\Theta - \Phi) \right\} \right] \nonumber
    \\
    &+& \frac{1}{6N_\text{triad}}\left[ Q(\Psi - \Phi) + R (\Phi^2 - \Phi) \right],
    \label{eq:Phi_ODE}
    \\
    \frac{d \Psi}{dt} &=&
    \frac{1}{6N_\text{tetrad}} R\left\{(1- \Psi) + 2(\Phi - \Psi) \right\} \nonumber\\
    &+&\frac{1}{6N_\text{tetrad}} 2(S_1 + S_2)\Phi(\Phi - 1). \label{eq:Psi_ODE}
\end{eqnarray}
Here $T$, $Q$, $R$, $S_1$, and $S_2$ are the numbers of sub-graphs on which the balance quantities are affected by an interaction between a pair of donors $d$ and $r$, i.e. $T$ for triads of $d$, $r$, and a third-party observer $o$ (Fig. \ref{fig:affected_subgraphs} (a)); $Q$ for tetrads of $d$, $r$, and third-party observers $o$ and $p$ (Fig. \ref{fig:affected_subgraphs} (b)); $R$ for trusses of $d$, $r$, and a third-party observer $o$, and a non-observing neighbor $n$ (Fig. \ref{fig:affected_subgraphs} (c)); $S_1$ for tetrads of $d$, a third-party observer $o$, and two non-observing neighbor $n$ and $m$ (Fig. \ref{fig:affected_subgraphs} (d)); and $S_2$ for tetrads of $d$, two third-party observer $o$ and $p$ , and a non-observing neighbor $n$ (Fig. \ref{fig:affected_subgraphs} (e)).
A truss is a four-node graph in which all but one pair of nodes are connected.
A non-observing neighbor for a donor-recipient pair is a node that are connected to the recipient and to a third-party observer but not to the donor.
These equations tell that non-observing neighbors are driving the system toward imbalance as sub-graphs with non-observing neighbors (Fig. \ref{fig:affected_subgraphs} (c), (d), (e)) always decrease triad and tetrad balance, i.e., the terms for $R$, $S_1$ and $S_2$ in Eqs. (\ref{eq:Phi_ODE}) and (\ref{eq:Psi_ODE}) are always negative.
This is because non-observing neighbors do not change the opinion while other observers may change their opinion about the donor, which is on average likely to make their triads or tetrads imbalanced.

For our present case of Er\"{o}des-R\'{e}nyi random graph with $N \gg 1$, the expected numbers of the subgraphs of a donor-recipient pair are
\begin{align}
  T &\simeq N p^2, \nonumber\\
  Q &\simeq \frac{1}{2} N^2p^5, \nonumber\\
  R &\simeq \frac{1}{2} N^2p^4(1-p), \nonumber\\
  S_1 &\simeq \frac{1}{2} N^3p^7(1-p)^2, \nonumber\\
  S_2 &\simeq \frac{1}{2} N^3p^8(1-p).
\end{align}
By solving $\frac{d \Theta}{dt} = \frac{d \Phi}{dt} = \frac{d \Psi}{dt} = 0$, the condition for $\Theta = \Phi = \Psi = 1$ to be the only stable fixed point of Eqs. (\ref{eq:Theta_ODE}), (\ref{eq:Phi_ODE}), and (\ref{eq:Psi_ODE}) is
\begin{equation}
 3\left[\left(\frac{2+T}{1+T}\right) - 3 \lambda \right] + Tp(1 - 2 \lambda) > 0,
\end{equation}
where $\lambda \equiv R/T = Np^2(1-p)$ is the average number of non-observing neighbors of each triad.
In the dense regime where $T \gg 1$, the second term dominates and hence the condition reduces to $\lambda < 1/2$.
In the sparse regime where $Tp \ll 1$, the first term dominates.
Since $(2+T)/(1+T)$ is bounded between $1$ and $2$, there is a value $2/3 \leq \lambda_{c} \leq 1$ such that $\lambda < \lambda_{c}$ makes the system to be absorbed.
This means that the number of non-observing neighbors should not exceed a certain value for the system to be in the absorbing phase.
These results from the mean-field analysis are consistent with the simulation results, in that
the system has the absorbing phase for the sparse and dense edge density regions, $Np^2(1-p) < 1$, while the active phase is in the middle of them.

The mean-field Eqs. (\ref{eq:Theta_ODE})-(\ref{eq:Psi_ODE}) provide us further understandings of the opinion dynamics, including the time scales of the relaxations of the order parameters (see Appendix in detail).
In the sparsest regime $Np = c \sim {\cal O}(1)$,
in which triangles are formed but the density is small to have percolated clusters of those,
the asymptotic forms of the equations tell us that
an autonomous relaxation of $\Theta$ to $1$ and the subsequent relaxation of $\Phi$ to $1$ take place fast. After these fast relaxations, the relatively slow relaxation of $\Psi$ takes place as follows 
\begin{equation}
\frac{d \Psi}{dt} = \frac{1-\Psi}{c N^3},
\end{equation}
meaning that $\Psi$ is also driven to an ordered value of $1$.

In the regime with more links: $Np^2 \sim {\cal O}(1)$, which corresponds to the point around the percolation of triangles, $\Theta$ and $\Phi$ relax to equilibrium values, which are $1$ if $c < \sqrt{2}$ and less than $1$ if $c > \sqrt{2}$. This process is followed by the slower and independent relaxation of $\Psi$ to $0$.
In more densely connected systems in which $Np = {\cal O}(N)$ with $p < 1 - {\cal O}(N^{-1})$, the asymptotic form of the dynamical equations yield the relaxation time of all the order parameters to be of the same order, and giving rise to a para-phase equilibrium $(\Theta_*, \Phi_*, \Psi_*) = (0, 0, 0)$.

In the most densely connected regime in which $1-p = c/N$, the relatively fast processes lead to relaxations of $\Theta_* \to \Phi$ and $\Phi_* \to \Psi$.
The dynamics of $\Psi$ after these fast relaxation follows 
\begin{equation}
    \frac{d \Psi}{dt}
    \sim
    \frac{(1- \Psi) [1 - (1 + c) \Psi]}{2N^2},
\end{equation}
which corresponds to the equilibrium
\begin{equation}
    \Theta_* = \Phi_* = \Psi_* = \frac{1}{1+c}.
\end{equation}
This again tells that the system shows a transition from the para-phase $(\Theta_*, \Phi_*, \Psi_*) = (0, 0, 0)$ to the ordered phase $(\Theta_*, \Phi_*, \Psi_*) = (1, 1, 1)$ at around $1-p \sim N^{-1}$, in the limit $N \to \infty$.

\section{Discussion}\label{sec:discussion}
In this study, through numerical simulations and mean-field analysis we found that, as a result of the indirect reciprocity, the density of social networks drastically changes the friendship and enmity structure.
In contrast with complete or fully connected networks~\cite{Oishi2013}, in which agents split into two fixed clusters and cooperate only within its own cluster, their relation (who likes or dislikes whom) keep changing in a wide range of network density.
The friendship and enmity structure was found to get fixed only if the network is sparse or dense enough, i.e., $p^2(1-p) < 1/N$.

A similar absorbing phase transition in friendship and enmity networks was observed in Heider's structural balance models~\cite{Heider1946, Cartwright1956, Antal2006, Radicchi2007, Radicchi2007a, Marvel2009, Marvel2011, Summers2013, Traag2013a, Nishi2014, Wongkaew2015}.
The structural balance models assume that agents either mutually like or dislike each other and change their opinion to increase their triad balance.
A structural balance model on random networks at certain edge densities shows a phase transition from absorbing phase in sparse networks to active phase in dense networks~\cite{Radicchi2007}.
In contrast, the present indirect reciprocity model allow agents to have different opinions of each other and assume the action and its observation as the reason behind the changes of their relation. 
This in turn result in changes of triads balance.
Furthermore, when we increase the edge density, the indirect reciprocity model shows an additional phase transition from active to absorbing phase for the higher density region, in addition to the absorbing-to-active transition in the lower density region like the structural balance model.

The transition in the lower density region can have a significant implication for a variety of complex networks. It is common that we study, either theoretically and empirically, sparse networks with average degree $k \sim \mathcal{O}(1)$, which corresponds to $p \sim \mathcal{O}(N^{-1})$ in our model. Therefore, the lower transition point $p_{*} \sim 1/\sqrt{N}$ is not unrealistically low and the transition can be relevant for various large networks such as online social networks~\cite{Leskovec2010, Szell2010b}.

Moreover, the transition in the higher density region also has negligible implications.
The value of the higher transition point $p^{*} \sim 1 - 1/{N}$ means that the agent networks need to be almost fully connected to be in the absorbing phase, in the sense that agents need to know all the other agents except for at most one agent.
For large networks, e.g., for those with millions of nodes, the requirement is likely too strict and the transition in the higher density is unlikely to be observed. However, important examples of signed or like--dislike social networks includes those of moderate sizes. For example, international relations has recently been analyzed as singed networks~\cite{Maoz2007, Warren2010, Doreian2015, Li2017} while the networks usually consist of some hundreds of nodes.
Another important example of signed social networks of moderate size and high density is who likes or dislikes whom in an organization (e.g., firms, schools)~\cite{Park2016}.

There are several questions yet to be answered in future studies. One is indirect reciprocity dynamics in real (or more realistic models of) social networks. Real social networks are not considered random in contrast to the present model and the network characteristics other than edge density such as degree distribution or community structure may largely affect the indirect reciprocity dynamics and friendship--enmity structure of our society. Yet another interesting issue is the character of the observed transition. Though beyond the scope of the present study, it is desirable to investigate whether the transitions are continuous or not, and if continuous whether they belong to the directed percolation university class~\cite{Hinrichsen2000,Lubeck2004}.

%

\appendix
\section{Updates of the edge balance}
In the following we denote the values of variable $x$ before and after an update at a time step as $x$ and $x'$, respectively.
Let $d$, $r$, and $o$ be the donor, the recipient, and a common neighbor of them at the time step (i.e. $d, r$, and $o$ form a triad).
Then only the following opinions are updated after the time step as
\begin{eqnarray}
    \sigma_{rd}' &=& \sigma_{dr}, \\
    \sigma_{od}' &=& \sigma_{or} \sigma_{dr}.
\end{eqnarray}
Therefore, the edge balance that can be updated to a different sign after the time step are those on $(d,r)$ and $(d,o)$ edges:
\begin{eqnarray}
  \Theta_{dr}' &=& \sigma'_{dr}\sigma'_{rd} = \sigma_{dr}^2 = 1, \nonumber \\
  \Theta_{do}' &=& \sigma'_{do}\sigma'_{od} = \sigma_{do} (\sigma_{or} \sigma_{dr}) = \Phi_{dor}.
  \label{eq_reciprocity}
\end{eqnarray}

\section{Updates of the triad balance}
Update rules of the triad balance are more complicated.
The first triad to be considered is the one formed by the donor $d$, the recipient $r$, and the observer $o$ (Fig. \ref{fig:affected_subgraphs} (A)).
While the triad balance $\Phi_{dro}$ and $\Phi_{dor}$ are kept under the opinion change,
the updated sign of other four quantities depends on the opinions before the update:
\begin{eqnarray}
    \Phi_{odr}'
    &=& \sigma'_{od}\sigma'_{or}\sigma'_{dr}\nonumber \\
    &=& (\sigma_{or}\sigma_{dr})\sigma_{or}\sigma_{dr} = 1,\nonumber \\
    \Phi_{ord}'
    &=& \sigma'_{or}\sigma'_{od}\sigma'_{rd}\nonumber \\
    &=& \sigma_{or}(\sigma_{or}\sigma_{dr})\sigma_{dr} = 1,\nonumber \\
    \Phi_{rod}'
    &=& \sigma'_{ro} \sigma'_{rd} \sigma'_{od}\nonumber \\
    &=& \sigma_{ro} \sigma_{dr} (\sigma_{or} \sigma_{dr})
    = \Theta_{ro}
    ,\nonumber \\
    \Phi_{rdo}'
    &=& \sigma'_{rd} \sigma'_{ro} \sigma'_{do}\nonumber \\
    &=& \sigma_{dr} \sigma_{ro} \sigma_{do}
    = \Phi_{dro}
    .
    \label{eq_triangle_DRO}
\end{eqnarray}
The second type of triads we should consider is the ones formed by the donor $d$ and the two observers of the time step knowing each other, $o$ and $p$ (Fig. \ref{fig:affected_subgraphs} (B)).
\begin{eqnarray}
  \Phi_{opd}'
  &=& \sigma'_{op} \sigma'_{od} \sigma'_{pd}\nonumber\\
  &=& \sigma_{op} (\sigma_{or}\sigma_{rd}) (\sigma_{pr}\sigma_{dr}) = \Phi_{opr},\nonumber \\
  \Phi_{odp}'
  &=& \sigma'_{od} \sigma'_{op} \sigma'_{dp}\nonumber\\
  &=& (\sigma_{or} \sigma_{dr}) \sigma_{op} \sigma_{dp} = \Psi_{odpr},\nonumber \\
  \Phi_{pod}' &=& \Phi_{por},\nonumber \\
  \Phi_{pdo}' &=& \Psi_{pdor}.
  \label{eq_triangle_OPD}
\end{eqnarray}
and because of the symmetry between $o$ and $p$,
  
The third and last triads one must take into account involve a non-observing neighbor $n$, who is connected to observer $o$ and the donor $d$ but not connected to the recipient $r$ (Fig. \ref{fig:affected_subgraphs} (C)). Because of the flip of $\sigma_{od}$ which takes place if $\Phi_{odr} = -1$, the following triad balance are updated as:
\begin{eqnarray}
    \Phi_{nod}' &=& \Phi_{odr} \Phi_{nod},
    \nonumber \\
    \Phi_{odn}' &=& \Phi_{odr} \Phi_{odn},
    \nonumber \\
    \Phi_{ond}' &=&  \Phi_{odr} \Phi_{ond}.
    \label{eq_triangle_OND}
\end{eqnarray}

\section{Updates of tetrad balance}
We have seen that the edge balance after a time step is determined by the triad balance, and the updates of triad balance can be descried by the edge, triad, and tetrad balance.
So we next consider the update rules of tetrad balance.
Updated tetrads are divided into those include the recipient and two third-party observers, and those include three third-party observers.
Note that all updated tetrads include the donor.
For the later type, we consider tetrads of $d, r$ and $o, p$ (Fig. \ref{fig:affected_subgraphs} (B)).
Then $\Psi_{opdr}$, $\Psi_{rpdo}$, and $\Psi_{rodp}$ are updated.
\begin{eqnarray}
  \Psi_{opdr}'
  &=& \sigma'_{od} \sigma'_{or} \sigma'_{pd} \sigma'_{pr}\nonumber\\
  &=& (\sigma_{or}\sigma_{dr})\sigma_{or} (\sigma_{pr}\sigma_{dr}) \sigma_{pr} \nonumber\\
  &=& (\sigma_{or}\sigma_{dr}\sigma_{pr})^2 = 1, \nonumber\\
  \Psi_{rpdo}'
  &=& \sigma'_{rd} \sigma'_{ro} \sigma'_{pd} \sigma'_{po}\nonumber\\
  &=& \sigma_{dr} \sigma_{ro} (\sigma_{pr}\sigma_{dr}) \sigma_{po}\nonumber\\
  &=& (\sigma_{pr}\sigma_{po}\sigma_{ro})\sigma^2_{dr} = \Phi_{pro},\nonumber\\
  \Psi_{rodp}'
  &=& \Phi_{opr}.
  \label{eq_quad_OPDR}
\end{eqnarray}
For the former type, we consider tetrads of $d$ and $o, p, q$ (Fig. \ref{fig:affected_subgraphs} (D)).
\begin{eqnarray}
  \Psi_{opqd}'
  &=& \sigma'_{oq} \sigma'_{od} \sigma'_{pq} \sigma'_{pd}\nonumber\\
  &=& \sigma_{oq} (\sigma_{or}\sigma_{dr}) \sigma_{pq} (\sigma_{pr}\sigma_{dr})\nonumber\\
  &=& (\sigma_{oq}\sigma_{or}\sigma_{pq}\sigma_{pr})\sigma_{dr}^2 = \Psi_{opqr},\nonumber\\
  \Psi_{pqod}' &=& \Psi_{pqor},\nonumber\\
  \Psi_{qopd}' &=& \Psi_{qopr}.
\end{eqnarray}

Then, we next take a mean-field treatment of the exact update rules considered above, by replacing the quantities in the right hand sides of the equations by those average at that time:
\begin{equation}
    \Theta_{ij} \ \sim \ \Theta,
    \
    \Phi_{ijk} \sim \Phi,
    \
    \Psi_{ijkl} \sim \Psi,
\end{equation}
which result in the Eqs. (\ref{eq:Theta_ODE})-(\ref{eq:Psi_ODE}).

\section{Asymptotic behavior of mean-field dynamics}
In the sparsest regime $Np = c \sim {\cal O}(1)$,
in which triangles are formed but the density is small to have percolated clusters of those, the Eqs. (\ref{eq:Theta_ODE})-(\ref{eq:Psi_ODE}) become to read as follows
\begin{eqnarray}
    \frac{d \Theta}{dt} 
    &\sim&
    \frac{c}{N} (1 - \Theta),
    \\
    \frac{d \Phi}{dt} 
    &\sim&
    \frac{c}{6N} \big[ (1 - \Phi) + (\Theta - \Phi) \big],
    \\
    \frac{d \Psi}{dt}
    &\sim&
    \frac{(\Theta -1) \Psi}{2N^2}
    +
    \frac{c \left[ (1- \Psi) + (\Phi - \Psi) \right]}{2N^3}.
\end{eqnarray}
This means that the autonomous relaxation of $\Theta$ to $1$ and the subsequent relaxation of $\Phi$ to $1$ take place fast.
And after reaching these equilibrium values, the leading term of the relatively slow relaxation of $\Psi$ becomes
\begin{equation}
\frac{d \Psi}{dt} = \frac{1-\Psi}{c N^3},
\end{equation}
meaning that $\Psi$ is also driven to an ordered value $1$.

In the regime with more links,
$Np^2 = c \sim {\cal O}(1) > 1$,
which corresponds to the point around the percolation of triangles,
the asymptotic dynamical equations reads as follows 
\begin{eqnarray}
    \frac{d \Theta}{dt} 
    &\sim&
    \frac{(1 - \Theta) + c (\Phi - \Theta)}{\sqrt{c} N^\frac{3}{2}},
    \\
    \frac{d \Phi}{dt} 
    &\sim&
    \frac{(1 + \Theta -2\Phi)
    - 3c \Phi (1 - \Phi)}{6 \sqrt{c} N^\frac{3}{2}},\\
    \frac{d \Psi}{dt}
    &\sim&
    - \frac{(1- \Theta)}{2N^2}\Psi.
\end{eqnarray}
The first two equations correspond to relatively fast dynamics, which has equilibrium at
$\Theta_* = \frac{2c+3}{(c+1)^2}, \ \Phi_* = \frac{c+2}{c(c+1)}$
other than the trivial one, $(\Theta_*, \ \Phi_*) = (1, 1)$.
Therefore these order parameters relaxes to an not-fully ordered value ($<1$) if $c > \sqrt{2}$.
The slower relaxation of $\Psi$ is a decay to $0$, though it can be very slow when $c < \sqrt{2}$ and hence $\Theta$ decays faster to the equilibrium $\Theta* = 1$.

In more densely connected systems in which $Np = {\cal O}(N)$ with
$p < 1 - {\cal O}(N^{-1})$, 
the asymptotic form of the dynamical equations are
\begin{eqnarray}
    \frac{d \Theta}{dt} 
    &\sim&
    \frac{p}{N}(\Phi - \Theta),
    \\
    \frac{d \Phi}{dt} 
    &\sim&
    \frac{p}{6N}
    \Big[ -(1-p) \Phi (1 - \Phi) + p (\Psi - \Phi) \Big],\\
    \frac{d \Psi}{dt}
    &\sim&
    - \left( \frac{p^2(1-p)}{2N} \right)
    \Psi (1 - \Psi).
\end{eqnarray}
This yields a para-phase equilibrium $(\Theta_*, \Phi_*, \Psi_*) = (0, 0, 0)$.

In the most densely connected regime in which $1-p = c/N$, 
the asymptotic form of dynamical equations are
\begin{eqnarray}
    \frac{d \Theta}{dt} 
    &\sim&
    \frac{\Phi - \Theta}{N},
    \\
    \frac{d \Phi}{dt} 
    &\sim&
    \frac{\Psi - \Phi}{6N},\\
    \frac{d \Psi}{dt}
    &\sim&
    \frac{(1- \Psi) (1 - c \Psi) - (1 - \Theta)\Psi}{2N^2}.
\end{eqnarray}
Under this dynamics, the relatively fast processes lead to relaxations of 
$\Theta_* \to \Phi$
and
$\Phi_* \to \Psi$.
Therefore, we can replace $\Theta$ in the last equation by $\Psi$, to obtain 
\begin{equation}
    \frac{d \Psi}{dt}
    \sim
    \frac{(1- \Psi) [1 - (1 + c) \Psi]}{2N^2}.
\end{equation}
From this we have the equilibrium
\begin{equation}
    \Theta_* = \Phi_* = \Psi_* = \frac{1}{1+c},
\end{equation}
which tells that this system shows a transition from the para-phase
$(\Theta_*, \Phi_*, \Psi_*) = (0, 0, 0)$
to the ordered phase
$(\Theta_*, \Phi_*, \Psi_*) = (1, 1, 1)$
at around $1-p \sim N^{-1}$, in the $N \to \infty$ limit.
Together with the results above, this system has an ordered state as the absorbing state in the regime
$p < {\cal O}(N^{-\frac{1}{2}})$
and
$1 - p < {\cal O}(N^{-1})$,
and otherwise it goes to a para-phase.

\end{document}